\documentclass[bibyear]{aa} 

\usepackage{graphicx}
\usepackage{txfonts}
\usepackage{hyperref}
\usepackage{float}

\usepackage{multirow}
\usepackage{natbib}
\bibpunct{(}{)}{;}{a}{}{,} 

\begin{document}

\title{A Gaia astrometric view of the open clusters Pleiades, Praesepe and Blanco 1}
\author{Jeison Alfonso
      \and
      Alejandro Garc\'{i}a-Varela
      }

\institute{Universidad de los Andes,
          Departamento de F\'{i}sica
          Cra. 1 No. 18A-10, Bloque Ip, A.A. 4976
          Bogot\'{a}, Colombia\\
          \email{[josegarc;je.alfonso1]@uniandes.edu.co}
          }
          
\date{Received date / Accepted date}

  \abstract
   {Near open clusters as Pleiades, Praesepe and Blanco 1 have been extensively studied due to their proximity to the Sun. The {\it Gaia} data brings the opportunity to investigate these clusters, since it contains valuable astrometric and photometric information which can be used to update their kinematic and stellar properties.}
   {Our goal is to carry out a star membership study in these nearby open clusters employing an astrometric model with proper motions and an unsupervised clustering machine learning algorithm using positions, proper motions and parallaxes. The star members are selected from the cross-matching between both methods. Once we know the members, we investigate the spatial distributions of these clusters and estimate their distances, ages and metallicities.}
   {We use the {\it Gaia} DR3 catalogue to determine star members using two approaches: a classical Bayesian model and the unsupervised machine learning algorithm DBSCAN. For star members we build the radial density profiles, the spatial distributions and compute the King parameters. The ages and metallicities were estimated using the BASE-9 Bayesian software.} 
   {We identified $958$, $744$ and $488$ star members for the Pleiades, Praesepe and Blanco 1 respectively. We corrected the distances and built the spatial distributions, finding that Praesepe and Blanco 1 have elongated shape structures. The distances, ages and metallicities obtained were consistent with the reported in the literature.}
   {We obtained catalogues of star members, updated kinematic and stellar parameters for these open clusters. We found that the proper motions model can find a similar number of members as the unsupervised clustering algorithm does when the cluster population form an overdensity in the vector point diagram. It allows to select an adequate size of the proper motions region to run these methods. Our analysis found stars that are being directed towards the outskirts of the Praesepe and Blanco 1, which exhibit elongated shapes. These stars have high membership probabilities and analogous proper motions as those within the tidal radius.}


\maketitle 

\section{Introduction}

Open clusters (OCs) are groups of stars, ranging from hundreds to thousands, held together by the gravitational force between their members. They have been found in spiral and irregular galaxies with active star formation processes \citep{reino2018}. OCs are formed from the collapse of molecular clouds \citep{krumholz2019} and have unique age and metallicity parameters. The distribution of proper motions and distances of cluster stars, resulting from the initial conditions of position, velocity, and angular momentum of the stars and their interaction with the Galaxy's potential over time, provide valuable information for studying stellar evolution.

The line of sight of OCs is often contaminated by background and foreground stars, making it challenging to determine which stars truly belong to the cluster. Nowadays, this star membership problem is a mandatory task to determine with highest precision the ages, metallicities, distances and rotational sequences of Milky Way OCs \citep{Godoy-Rivera2021}. To solve this issue, \citet{vasilevskis1958} modeled the proper motion distribution of cluster and field stars using circular and elliptical bivariate density functions respectively. By applying the Bayes rule, the probabilities of a star belonging to the cluster can be estimated based on its proper motions $(\mu_{\alpha}, \mu_{\delta})$. The proper motions model (hereafter PM model) is effective in identifying stellar members, but becomes problematic when the cluster is far away and numerical issues arise in finding the proper motion centroid of the cluster. In such cases, the computed membership probabilities may not be accurate. However, for nearby OCs their larger proper motions compared to those of field stars, make it easier to identify members using the Vector Point Diagram (VPD). This is the case of Pleiades, Praesepe, and Blanco 1 OCs, whose proper motions have a clear separation from the background stars. 

The Density-Based Spatial Clustering of Applications with Noise (DBSCAN) algorithm is another method used to identify the stars that belong to an OC \citep{ester1996}. This unsupervised machine learning algorithm detects overdensities in a multi-dimensional space to find clusters. The definition of clusters is based on two hyper-parameters: the $\epsilon$ radius, which defines the size of a neighborhood around a point, and the minimum number of points $(minPts)$ that a cluster must contain within the neighborhood defined by the $\epsilon$ radius. By using the {\it Gaia} Data Release 3 ({\it Gaia} DR3) positions, proper motions and parallaxes, and applying the DBSCAN algorithm, it is possible to search for OCs in a specific sky area. 

In this work, we aim to identify star members of Pleiades, Praesepe, and Blanco 1, using the Gaia DR3 catalogue. We employ both, the PM model, which estimates star membership probabilities using Maximum Likelihood Estimation (MLE) and Markov Chain Monte Carlo (MCMC) algorithm, and DBSCAN. We select the star members based on the cross-matching of the PM model and DBSCAN results. The distances are then corrected using the \citet{bailer2015} procedure. The structures of the clusters are analyzed by building Radial Density Profiles (RDPs) and spatial distributions, computing core and tidal radii from the King model fit, and estimating their ages and metallicities using Color-Magnitude Diagrams (CMDs) and PARSEC isochrones \citep{bressan2012,marigo2013}.

The paper is structured into five sections. Section \ref{sec:methods} is divided into three subsections that describe the PM model implementation, the DBSCAN clustering algorithm, the estimation of parameters, and the density distribution used to compute the King model. Section \ref{sec:data} describes the procedure to select the region of each cluster based on their average proper motions and distances. Section \ref{sec:results_and_discussion} is divided into five subsections, two of which describe the results of star membership in each cluster and their implementation. The last three describe the RDPs with the estimation of King parameters, the spatial distributions and the estimation of age and metallicity for each OC. The conclusions are presented in Sect. \ref{sec:conclusions}. Appendixes related to the {\it Gaia} data selection and the uncertainty estimation of parameters are presented.


\section{Methods}
\label{sec:methods}

\subsection{The proper motions model}
\label{sec:pm_model}

The segregation problem in OCs involves identifying cluster and field stars, which overlap in the VPD due to their relative positions. Cluster stars are grouped around the center of mass due to their gravitational interaction, while field stars are scattered in the foreground and background with weaker gravitational ties. One of the earliest studies on this problem was performed by \citet{vasilevskis1958} using proper motions obtained from photographic plates to determine the probable members of the NGC 6633 cluster. To distinguish between the two populations, these authors model the proper motion distribution by two bivariate probability densities: a circular density for the cluster stars described by Eq. \eqref{equ:cluster}, and an elliptical density for the field stars described by Eqs. \eqref{equ:field}, \eqref{equ:elliptic} (henceforth $c$ and $f$ subscript for cluster and field). By combining these two probability densities, the joint probability distribution can be obtained, as expressed by Eq. \eqref{equ:join_density}, \citep{sanders1971,slovak1977}.

\begin{equation}\label{equ:cluster}
    \psi_{c}(\mu_{\alpha i},\mu_{\delta i})=\frac{1}{2 \pi \sigma_c^{2}} exp \left (-\frac{1}{2}\left [ \left (\frac{\mu_{\alpha i}-\mu_{\alpha c}}{\sigma_c}\right)^{2} +\left ( \frac{\mu_{\delta i}-\mu_{\delta c}}{\sigma_c}\right)^{2}\right] \right ),
\end{equation}
\vspace{-0.5cm}
\begin{equation}\label{equ:field}
    \psi_{f}(\mu_{\alpha i},\mu_{\delta i})=\frac{1}{2 \pi \sigma_{\alpha f} \sigma_{\delta f} \sqrt{1-\rho^{2}}}exp\left [-\frac{1}{2(1-\rho^{2})} \Omega (\mu_{\alpha i},\mu_{\delta i}) \right ],
\end{equation}

with

\begin{multline}\label{equ:elliptic}
 \Omega (\mu_{\alpha i},\mu_{\delta i}) =  \left (\frac{\mu_{\alpha i}-\mu_{\alpha f}}{\sigma_{\alpha f}} \right )^{2} + \left (\frac{\mu_{\delta i}-\mu_{\delta f}}{\sigma_{\delta f}} \right)^{2} \\
 -2 \rho \left (\frac{\mu_{\alpha i}-\mu_{\alpha f}}{\sigma_{\alpha f}} \right )\left (\frac{\mu_{\delta i}-\mu_{\delta f}}{\sigma_{\delta f}} \right ),
\end{multline}

\begin{equation}\label{equ:join_density}
    \psi(\mu_{\alpha i}, \mu_{\delta i}) = n_c \psi_{c}(\mu_{\alpha i}, \mu_{\delta i}) + (1-n_c)\psi_{f}(\mu_{\alpha i}, \mu_{\delta i}).
\end{equation}

The joint probability distribution in Eq. \eqref{equ:join_density}, is a function of nine parameters represented by a $\vec{\theta}$ vector. It includes the centroids of proper motion distributions for the cluster and field stars ($\mu_{\alpha c}$, $\mu_{\delta c}$, $\mu_{\alpha f}$, $\mu_{\delta f}$), the standard deviations ($\sigma_c$, $\sigma_{\alpha f}$, $\sigma_{\delta f}$), the correlation coefficient ($\rho$) between proper motions, and the fraction of cluster stars ($n_c$) relative to the total number of stars ($N = N_c + N_f$). To compute these parameters we employed the MLE. This method uses the log-likelihood function in Eq. \eqref{equ:likelihood}, defined from the joint probability distribution and gives the probability of the proper motions sample in the VPD as a function of the parameters \citep{uribe1994}.

\begin{equation}\label{equ:likelihood}
    \ln\mathcal{L}(\vec{\theta}) =  \sum_{i=1}^n \ln \psi(\mu_{\alpha i}, \mu_{\delta i}| \vec{\theta}).
\end{equation} 

The log-likelihood function predicts the observed proper motions based on the parameters, which are obtained through the implementation of the MLE. The method maximizes the log-likelihood function to find the nine parameters that generated the observed data. The initial guess was derived from the marginal densities provided by \citet{sabogal2001}. To calculate the membership probabilities of each star, the bivariate probability densities given by Eqs. \eqref{equ:cluster}, \eqref{equ:field}; are used in accordance with Bayes theorem. The apriori probabilities of belonging to the cluster ($n_c$) or the field ($n_f$) are established for each population \citep{uribe1994}. Finally, the membership probability of the $i_{th}$ star with proper motions $\mu_{\alpha i}$ and $\mu_{\delta i}$, is given by Eq. \eqref{equ:membership}.

\begin{equation}\label{equ:membership}
    P_i (cluster | \mu_{\alpha i}, \mu_{\delta i}) = \frac{n_c \psi_{c}(\mu_{\alpha i},\mu_{\delta i})}{n_c \psi_{c}(\mu_{\alpha i},\mu_{\delta i}) + (1-n_c) \psi_{f}(\mu_{\alpha i},\mu_{\delta i})}.
\end{equation}

To determine if a star belongs to the cluster, we establish a threshold of $P \geq 0.5$. If the membership probability of a star is greater than or equal to $0.5$, it is considered to be part of the cluster.


\subsection{The DBSCAN algorithm}
\label{sec:dbscan}

The recent use of the unsupervised clustering algorithm DBSCAN by \citet{castro-ginard2018} with the {\it Gaia} DR2 catalogue involves identifying clusters through positions, proper motions and parallaxes. The algorithm identifies overdensities by computing distances between points in the data set. DBSCAN is dependent on two parameters: the $\epsilon$ radius of the hypersphere and the minimum number of points ($minPts$) that must be inside that hypersphere \citep{ester1996}. In this algorithm each point in the data set is assigned to one of the following three categories: (1) core points that contain at least $minPts$ number of points, including the point itself, within their $\epsilon$ radius surrounding area in the dimensional space, (2) border points that are reachable from a core point and have less than $minPts$ number of points within their $\epsilon$ radius surrounding area, and (3) outliers, which are points that are neither core nor border points.

To employ the clustering algorithm, we adopt a five-dimensional space composed of the equatorial coordinates, proper motions and parallaxes. We utilize the standard euclidean metric as expressed in Eq. \eqref{equ:metric}, to measure the distances between the $i$ and $j$ stars.

\tiny
\begin{equation}\label{equ:metric}
    d(i,j)=\sqrt{(\alpha_{i}-\alpha_{j})^2+(\delta_i - \delta_j)^2+(\mu_{\alpha i}-\mu_{\alpha j})^2+(\mu_{\delta i}-\mu_{\delta j})^2 + (\varpi_i-\varpi_j)^2},
\end{equation}
\normalsize

where the data has been previously standardised.


\subsection*{Determination of the $\epsilon$ and $minPts$ parameters}

To optimize the results of the DBSCAN algorithm, an appropriate value of the $\epsilon$ parameter is essential. A procedure based on the one described by \citet{castro-ginard2018} was followed to determine an optimal $\epsilon$. The first step was to compute the $k_{th}$ Nearest-Neighbour Distance ($k$-NND) histogram of the data, and the median value was stored as $\epsilon_{kNN}$. Next, a random sample of the same size as the data was generated using a Metropolis-Hastings algorithm from a Gaussian kernel. Then, the $k$-NND histogram of the simulated data was computed, and the median value was stored as $\epsilon_{rand *}$. To reduce variation in $\epsilon_{rand *}$ due to sampling, an average of $30$ repetitions was performed, and the result was stored as $\epsilon_{rand}$. The final value of $\epsilon$ was determined as the average of $\epsilon_{rand}$ and $\epsilon_{kNN}$, $\epsilon = (\epsilon_{rand}+\epsilon_{kNN})/2$.

\begin{figure}[htp]
    \centering
    \includegraphics[scale=0.78]{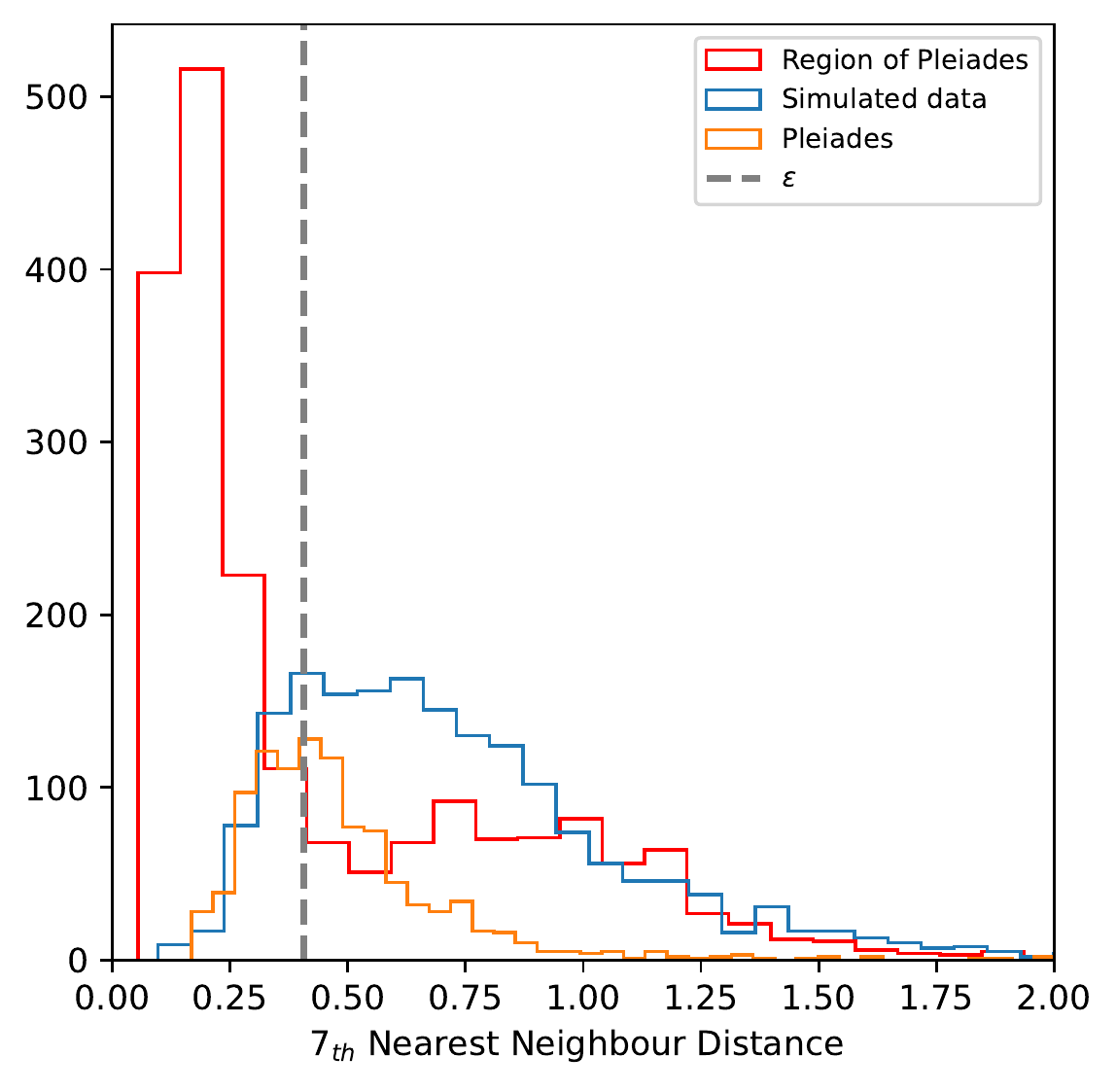}
    \caption{Histogram of the $7_{th}$-NND for the region of Pleiades (red), simulated data (blue) and the Pleiades members selected from the PM model (orange).
    The gray dashed vertical line is the selected $\epsilon$.}
    \label{fig:knnd}
\end{figure}

The distribution of the $7_{th}$-NND for the Pleiades cluster region is shown in Fig. \ref{fig:knnd}. The graph displays the $7_{th}$-NND for the Pleiades members selected from the PM model, the simulated data generated from a Gaussian kernel, and the selected $\epsilon$ value in that region. Unlike the approach used by \citet{castro-ginard2018} where the minimum value of the $k$-NND for the cluster and simulated data was chosen as the average, the distribution of the members and simulated data in the Pleiades cluster were found to be similar. Therefore, we choose the median value to improve the efficiency of the algorithm. Also, selecting the minimum value as \citet{castro-ginard2018} did, would result in classifying all data as outliers for any value of $minPts$.

To determine the $minPts$ parameter in the algorithm, the size of the region being considered and the expected number of cluster members should be taken into account. \citet{castro-ginard2018} found that a range of $minPts$ between 5 and 9 provides a good balance between minimizing false positives and achieving good efficiency. 


\begin{figure*}
\centering
\includegraphics[width=0.3\textwidth]{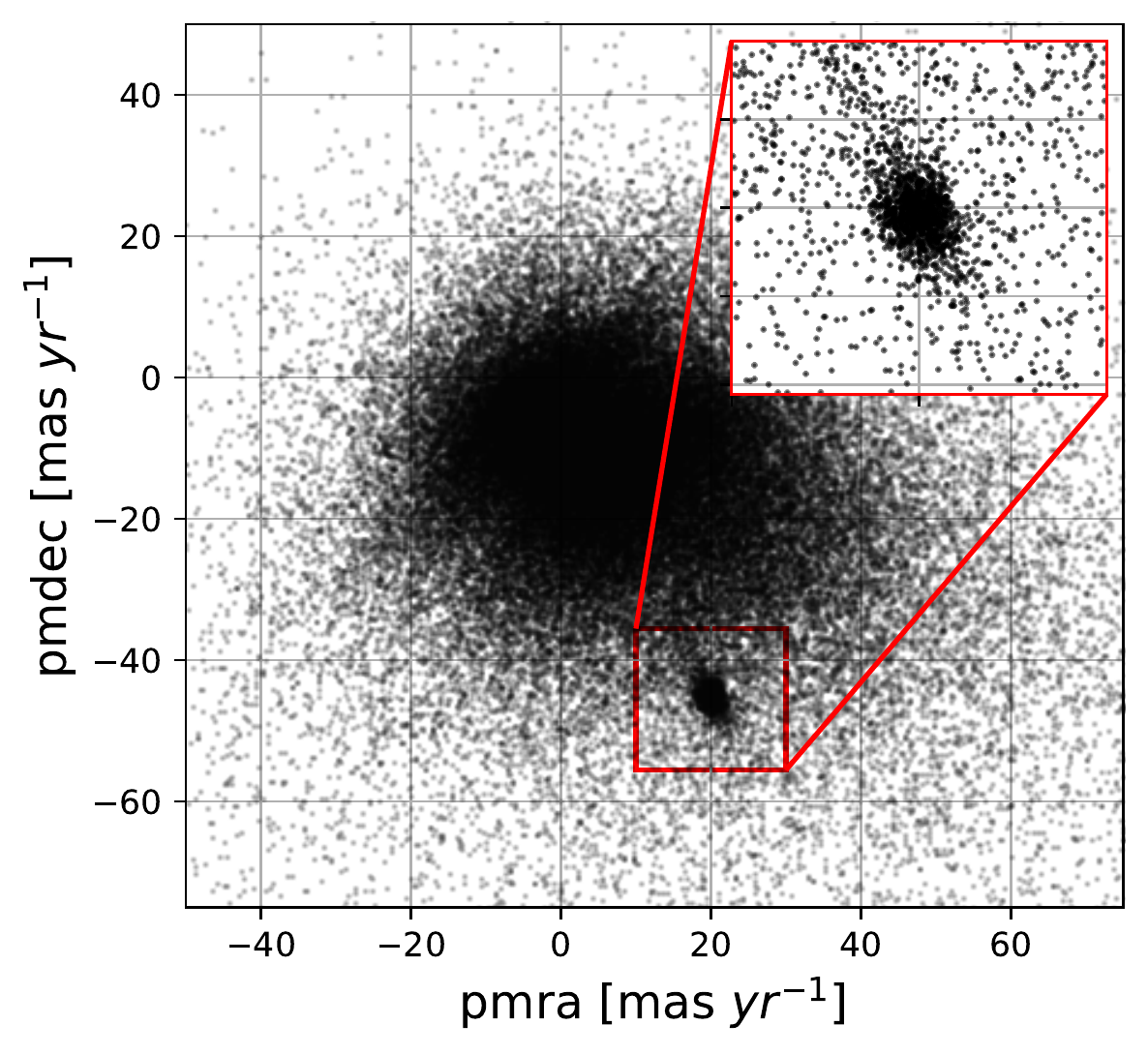}
\includegraphics[width=0.3\textwidth]{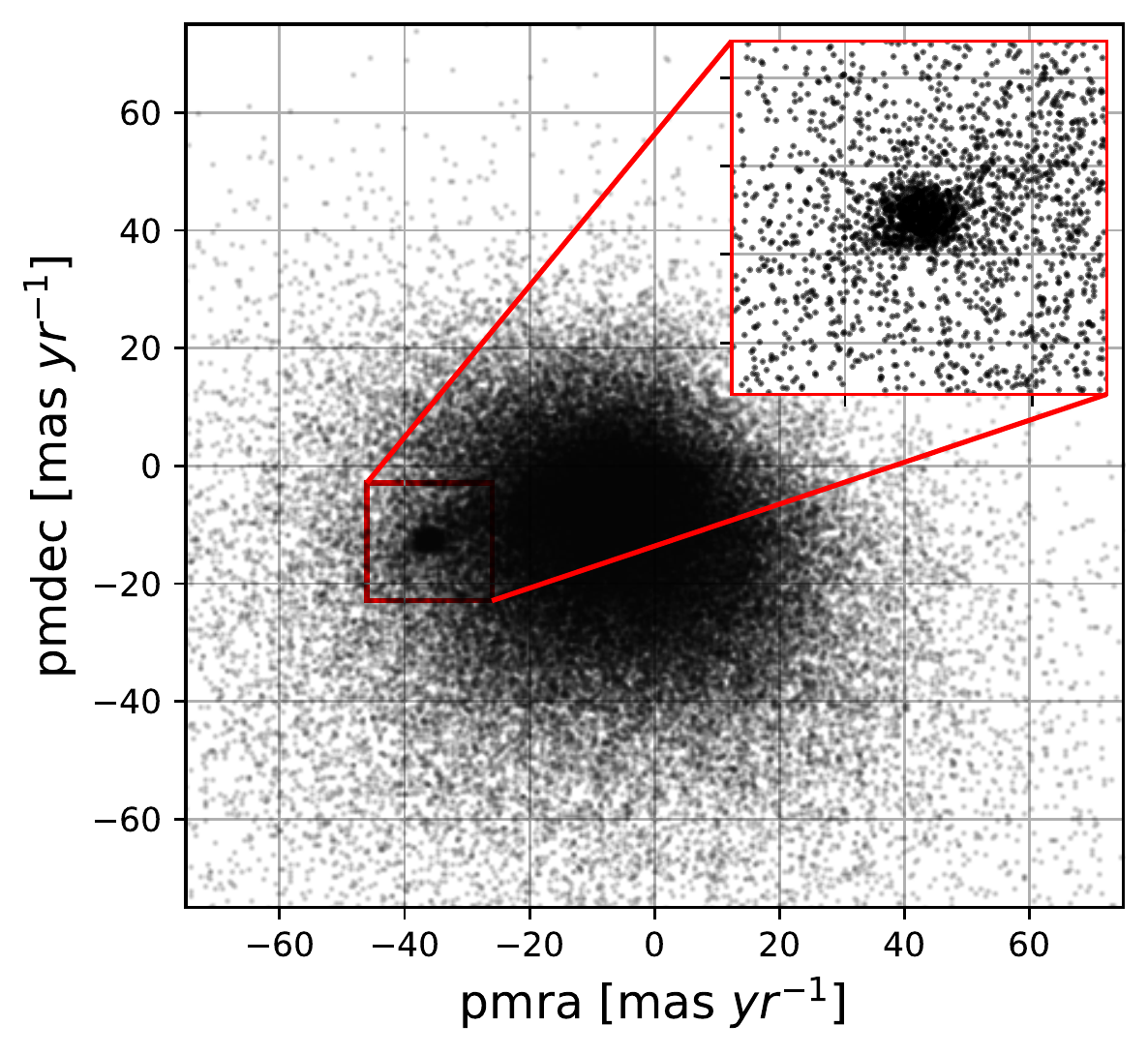}
\includegraphics[width=0.3\textwidth]{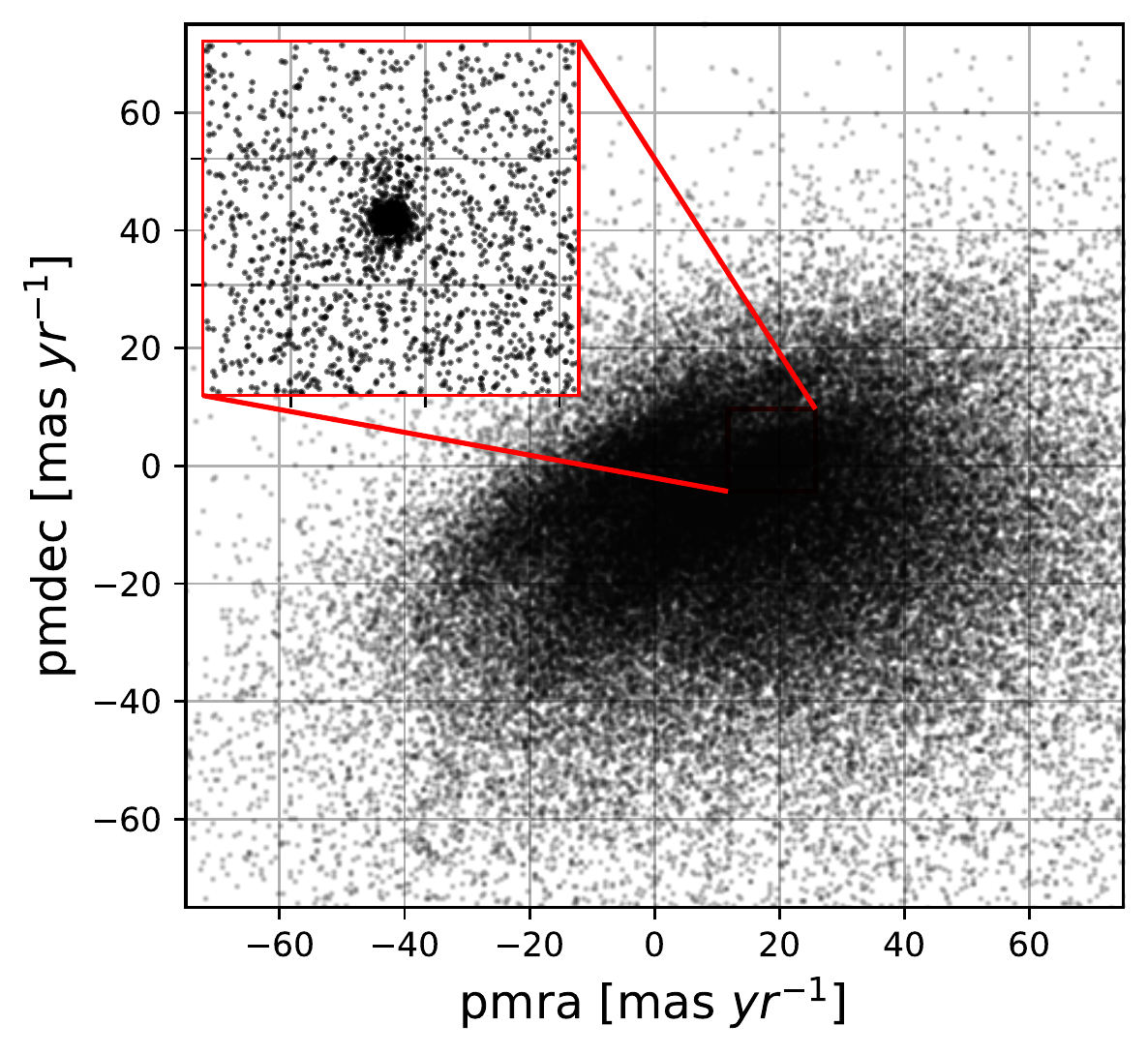}
\caption{VPDs for the Pleiades (left panel), Praesepe (middle panel) and Blanco 1 (right panel). The zoom plots show the selected regions to compute star membership with the PM model and the DBSCAN clustering algorithm.}\label{fig:vpd_zoom}
\end{figure*}


\subsection{The density distribution}
\label{sec:density_distribution}

The analysis of the structure of an OC could profit from the calculation of its stellar surface density. It provides information on the population of stars as a function of their projected distance from the center of the cluster. To model this distribution, the empirical surface density function introduced by \citet{king1962} is often used. This model was derived from the study of globular clusters in the Milky Way.

The King model accounts for both the inner and outer densities of the cluster and is described by two equations: $\rho = \rho_0/(1+(r/r_c)^2)$ for the inner and $\rho = \rho_1 ( 1/r - 1/r_t)^2$ for the outer density. The normalization constants $\rho_0$ and $\rho_1$ are used to scale the density values. $r_c$ is the core radius, the distance from the center in which the density falls to half its central value ($\rho_0$). On the other hand, $r_t$, the tidal radius, is the distance in which the cluster is tidally scattered by the Galaxy potential and the density reaches zero. These two equations can be condensed into a single expression, Eq. \eqref{equ:king_function}, to summarize the King model \citep{pinfield1998}.

\begin{equation}\label{equ:king_function}
    \rho(r) = \rho_0 \left[ \frac{1}{\sqrt{1 + (r/r_c)^2}} - \frac{1}{\sqrt{1 + (r_t/r_c)^2}} \right]^2,
\end{equation}

where $\rho_0$, $r_c$ and $r_t$ are the same as those described above. Even though the King model is not the physical solution of the collisionless Boltzmann equation, it is very useful to characterize the density profiles of star clusters \citep{binney2008}.


\section{Data}
\label{sec:data}

The European Space Agency (ESA) published the {\it Gaia} DR3 catalogue, which contains information on 1.8 billion sources collected over the first 34 months of the mission. The DR3 includes photometry in the $G$, $G_{BP}$, $G_{RP}$ pass-bands, radial velocities, and astrometric parameters including positions, proper motions and parallaxes \citep{gaia2016,gaiadr32022,babusiaux2022}. The data of the clusters under studied was downloaded from the {\it Gaia} archive using their centers in equatorial coordinates from SIMBAD and a search radius of 10 degrees. We used astrometric and photometric errors filtered out to eliminate spurious sources as recommended by \citet{lindegren2021} (see Appendix \ref{appendix:gaia_query}). These filters constrain the population of faintest stars. Nevertheless, further research is required to extend the cluster members using methodologies such as the one proposed by \citet{groeningen2023} and thus extend the cluster membership lists. Our search was also limited to 500 pc through a parallax cutoff.

The clusters being studied are located in close proximity to the Sun. Their proper motions reported by \citet{babusiaux2018} are centered at $(19.9 , - 45.5)$ mas yr$^{-1}$,s $(-36.1 , - 12.9)$ mas yr$^{-1}$ and $(18.7 , 2.6)$ mas yr$^{-1}$ for the Pleiades, Praesepe and Blanco 1, respectively. This leads to the formation of overdensities in the VPDs due to their greater changes in coordinates when compared to distant stars. Consequently, it becomes possible to remove part of the background stars and reduce data contamination. However, finding the cluster and field centroids using the PM model can be limited when $n_f$ significantly exceeds $n_c$. To address this, we fitted Gaussian densities around the cluster's average proper motions in each axis and choose stars within $3 \sigma$ range to select a region in order to study the star membership. Around a third of the stars outlined in the zoom plots in Fig. \ref{fig:vpd_zoom} have radial velocities reported.


\section{Results and Discussion}
\label{sec:results_and_discussion}

In this study, we aim to perform the star membership in three nearby OCs using data from the {\it Gaia} DR3 catalogue. This is achieved through two methods: the astrometric PM model and an unsupervised clustering algorithm DBSCAN. The PM model is based on the Bayesian approach and uses proper motions for membership determination \citep{vasilevskis1958}. On the other hand, the DBSCAN algorithm takes into consideration five dimensions: equatorial coordinates, proper motions and parallaxes \citep{ester1996,castro-ginard2018}. The final membership is determined by cross-matching the results from both methods\footnote{The full list of the star members considered in each cluster will be made available as online material.}.


\subsection{Star membership with the PM model}
\label{Membership_determination_with_proper_motions}

The PM model was implemented using two bivariate probability densities given by Eqs. \eqref{equ:cluster}, \eqref{equ:field}. The MLE method was employed to determine the nine components of the vector parameter $\Vec{\theta}$. The estimation was performed by utilizing the log-likelihood function given by Eq. \eqref{equ:likelihood}, which maps the vector parameter space $\vec{\theta}$ and gives a probability for the data. Subsequently, the vector parameter that maximizes the probability for the observed data can be obtained, resulting in the maximum likelihood estimate $\vec{\theta_{max}}$.

Aiming to determine the nine parameters of the PM model, we fitted a marginal density to the individual proper motion histograms. This tuning process resulted in the calculation of standard deviations and centroids for both the cluster and field stars, which served as the initial guess for the MLE method. Following, the maximization of the log-likelihood function was implemented considering the parameters previously determined as the initial guess. Additionally, in order to estimate the uncertainty of the nine parameters, a MCMC algorithm was performed \citep{foreman2013} (see the corner plot of the posterior probability function in the Pleiades region in Appendix \ref{appendix:mcmc}), utilizing an uniform prior and computing the upper and lower error for each parameter as specified in Table \ref{tab:parameters}. The convergence of the chains was validated by the \citet{gelman1992} diagnostic, we found $0.9994$, $0.9996$ and $1.0000$ $\hat{R}$ values for Pleiades, Praesepe and Blanco 1, respectively. It means that the convergence has been reached.

The computed fraction of cluster star values ($n_c$) were enough to offset the number of stars in the two populations, ensuring accurate membership probabilities. The correlation coefficients ($\rho$) were close to zero, indicating a lack of linear relationship between the proper motion distributions of cluster and field stars. The standard deviations of field stars were widely distributed compared to the cluster stars, as expected. Furthermore, the selected regions depicted in Fig. \ref{fig:vpd_zoom}, allowed to estimate the centroids of both cluster and field stars properly. The PM model estimates probabilities in a two-dimensional space, therefore it does not have the depth that other methods do. However, it can find stars exhibiting proper motions similar to the cluster mean, which may be helpful for observing the outskirts.

\renewcommand{\arraystretch}{1.3}
\begin{table}[H]
    \caption{The PM model parameters $\Vec{\theta_{max}}$ computed using the MLE method for Pleiades, Praesepe and Blanco 1 clusters. The upper and lower uncertainty in each parameter corresponds to the $16_{th}$ and $84_{th}$ percentiles of the samples obtained through the MCMC algorithm, respectively.}
    \label{tab:parameters}
    \centering
    \begin{tabular}{cccccc}\hline\hline
    \multirow{2}{*}{Parameter} & \multirow{2}{*}{Pleiades} & \multirow{2}{*}{Praesepe} & \multirow{2}{*}{Blanco 1} \\
    &  &  &  \\\hline
         $n_c$ & $0.48_{-0.01}^{+0.01}$ & $0.30_{-0.01}^{+0.01}$ & $0.27_{-0.01}^{+0.01}$ \\
         $\sigma_c$ & $1.11_{-0.03}^{+0.03}$ & $0.99_{-0.03}^{+0.03}$ & $0.39_{-0.01}^{+0.01}$ \\
         $\sigma_{\alpha f}$ & $4.95_{-0.11}^{+0.12}$ & $5.21_{-0.09}^{+0.10}$ & $3.75_{-0.07}^{+0.07}$ \\
         $\sigma_{\delta f}$ & $5.23_{-0.12}^{+0.12}$ & $5.00_{-0.09}^{+0.09}$ & $3.71_{-0.07}^{+0.07}$ \\
         $\rho$ & $-0.14_{-0.03}^{+0.03}$ & $0.08_{-0.02}^{+0.02}$ & $-0.02_{-0.03}^{+0.03}$ \\
         $\mu_{\alpha c}$ & $19.95_{-0.04}^{+0.04}$ & $-35.98_{-0.05}^{+0.05}$ & $18.73_{-0.02}^{+0.02}$ \\
         $\mu_{\delta c}$ & $-45.41_{-0.04}^{+0.05}$ & $-12.86_{-0.05}^{+0.05}$ & $2.58_{-0.02}^{+0.02}$ \\
         $\mu_{\alpha f}$ & $19.82_{-0.15}^{+0.16}$ & $-34.79_{-0.13}^{+0.13}$ & $18.99_{-0.10}^{+0.10}$ \\
         $\mu_{\delta f}$ & $-43.91_{-0.16}^{+0.17}$ & $-12.91_{-0.13}^{+0.12}$ & $1.91_{-0.10}^{+0.10}$ \\\hline
    \end{tabular}
\end{table}
\renewcommand{\arraystretch}{1.3}
\begin{table*}
\caption{Number of members found with the PM model, the DBSCAN (DB) algorithm, the Cross-matching and parameters computed of the target clusters.}
\label{tab:members}
\centering
\begin{tabular}{@{\extracolsep{4pt}}cccccccccc@{}}\hline\hline
\multirow{2}{*}{Cluster} & \multirow{2}{*}{PM} & \multirow{2}{*}{DB} & \multirow{2}{*}{Cross-matching} & $d$ & $d_L$ & logAge & logAge$_L$ & $Z$ & $Z_L$ \\ \cline{5-6} \cline{7-8} \cline{9-10}
&  &  &  & \multicolumn{2}{c}{(pc)} & \multicolumn{2}{c}{(Myr)} & \multicolumn{2}{c}{(dex)}  \\\hline 
Pleiades & $1041$ & $1026$ & $958$ & $135.74 \pm 0.10$ & $135.15 \pm 0.43$\tablefootmark{(a)} & $7.99$ & $8.04$\tablefootmark{(b)} & $0.015$ & $0.017$\tablefootmark{(b)} \\
Praesepe & $795$ & $885$ & $744$ & $184.72 \pm 0.19$ & $187.35 \pm 3.89$\tablefootmark{(a)} & $8.88$ & $8.87$\tablefootmark{(c)} & $0.020$ & $0.020$\tablefootmark{(b)} \\
Blanco 1 & $545$ & $514$ & $488$ & $236.67 \pm 0.25$ & $236.70 \pm 2.10$\tablefootmark{(d)} & $8.01$ & $8.06$\tablefootmark{(b)} & $0.017$ & $0.017$\tablefootmark{(b)} \\\hline
\end{tabular}
\tablefoot{The distance ($d$), logAge and metallicity ($Z$) are the values found in this work and those with $_L$ subscript are reported by the literature. \tablefoottext{a}{\citet{lodieu2019};} \tablefoottext{b}{\citet{babusiaux2018};} \tablefoottext{c}{\citet{gossage2018};} \tablefoottext{d}{\citet{pang2021}.}}
\end{table*}

The membership probabilities were calculated according to Eq. \eqref{equ:membership}, using the estimated parameters $\vec{\theta_{max}}$ reported in Table \ref{tab:parameters}. A star was classified as a member of the cluster if its membership probability was greater than or equal to a predetermined threshold value of $0.5$. The number of members identified by the PM model is presented in Table \ref{tab:members}.


\subsection{Star membership with the DBSCAN algorithm}

The classification of OCs was also performed using the DBSCAN clustering algorithm \citep{ester1996} in a five-dimensional astrometric space ($\alpha$, $\delta$, $\mu_{\alpha}$, $\mu_{\delta}$, $\varpi$). We do not include the radial velocities as an additional dimension because only bright stars have this measurement in the catalogue ($G_{RVS} \lessapprox 14$ mag). The distances between standardized data were calculated using the metric given by Eq. \eqref{equ:metric}. Our approach followed a similar methodology established by \citet{castro-ginard2018} to determine the parameters $minPts$ and $\epsilon$. In accordance with \citet{castro-ginard2018}, a reliable range for the value of $minPts$ was determined to be $5$ to $9$, based on the size of the region being studied and by minimizing the number of false positives in the clusters classified through the algorithm. We have selected $minPts = 8$ and the corresponding equation $k = minPts - 1$ was used to calculate the $k$-NND. We have also found that choosing a different value within the $minPts$ range would not significantly impact the number of star members identified in the clusters by DBSCAN.

The methodology employed for determining the parameter $\epsilon$ in DBSCAN deviates moderately to some extent from that of \citet{castro-ginard2018}. This is due to the fact that the original methodology results in the classification of all data points as outliers. To address this issue, we have determined a reliable value of $\epsilon$ by calculating the average of the $7_{th}$-NND between the median value of the data and the average of $30$ random resamplings, as described in Sect. \ref{sec:dbscan}. The values of $\epsilon$ determined for the Pleiades, Praesepe, and Blanco 1 regions are $0.404128$, $0.370619$ and $0.373202$, respectively. These values remained almost unchanged even when the resampling was repeated several times. Figure \ref{fig:knnd} depicts the $7_{th}$-NND histograms for the entire Pleiades region, the simulated data from a Gaussian kernel, and the Pleiades members selected from the PM model. The distributions of the cluster and simulated distances are in a similar range as the Pleiades members selected by the PM model, therefore the median value was used instead of the minimum, as was done by \citet{castro-ginard2018}. The algorithm is based on finding overdensities in the dimensional space, thus tending to gather stars which minimize the population of stars in the surroundings. 
Despite the widespread use of HDBSCAN detecting clusters in the {\it Gaia} catalogue \citep{mcinnes2017}, we choose DBSCAN since this algorithm allows to fine-tune the $\epsilon$ through sampling on the $k$-NND in the cluster regions.
The number of members identified by the DBSCAN algorithm is presented in Table \ref{tab:members}.


\subsection{Radial density profile}
\label{sec:rdp}

\begin{figure*}
\centering
\includegraphics[width=0.3\textwidth]{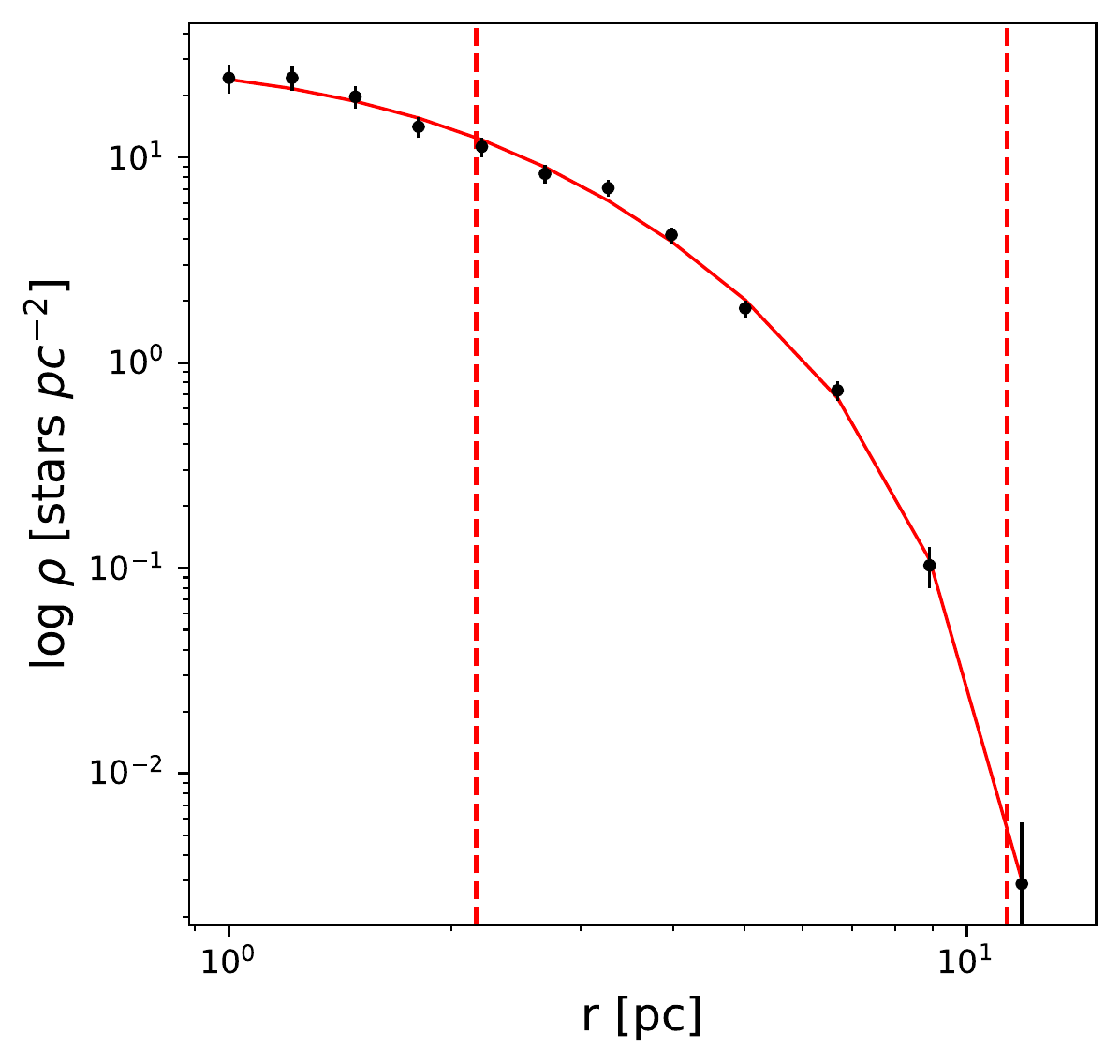}
\includegraphics[width=0.3\textwidth]{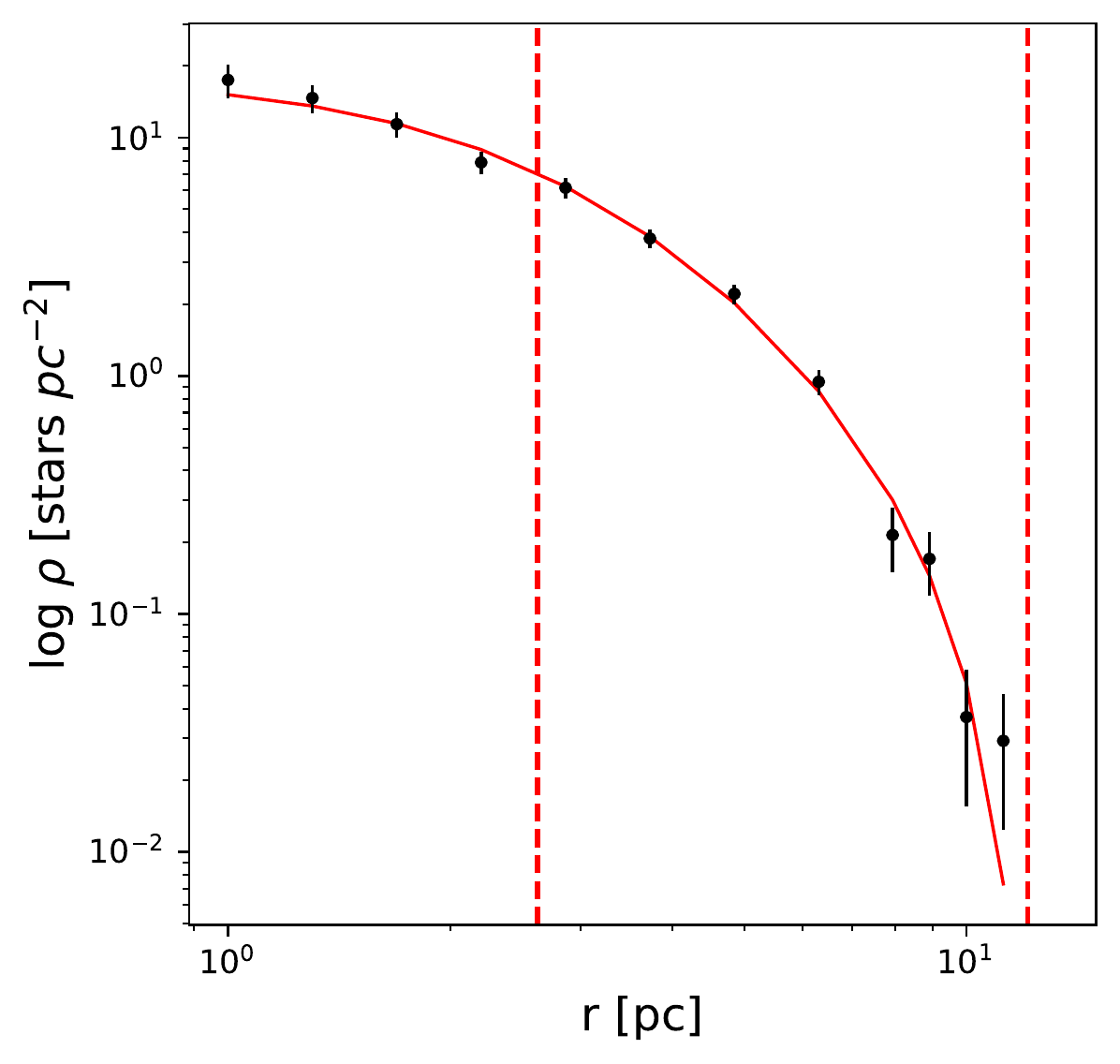}
\includegraphics[width=0.3\textwidth]{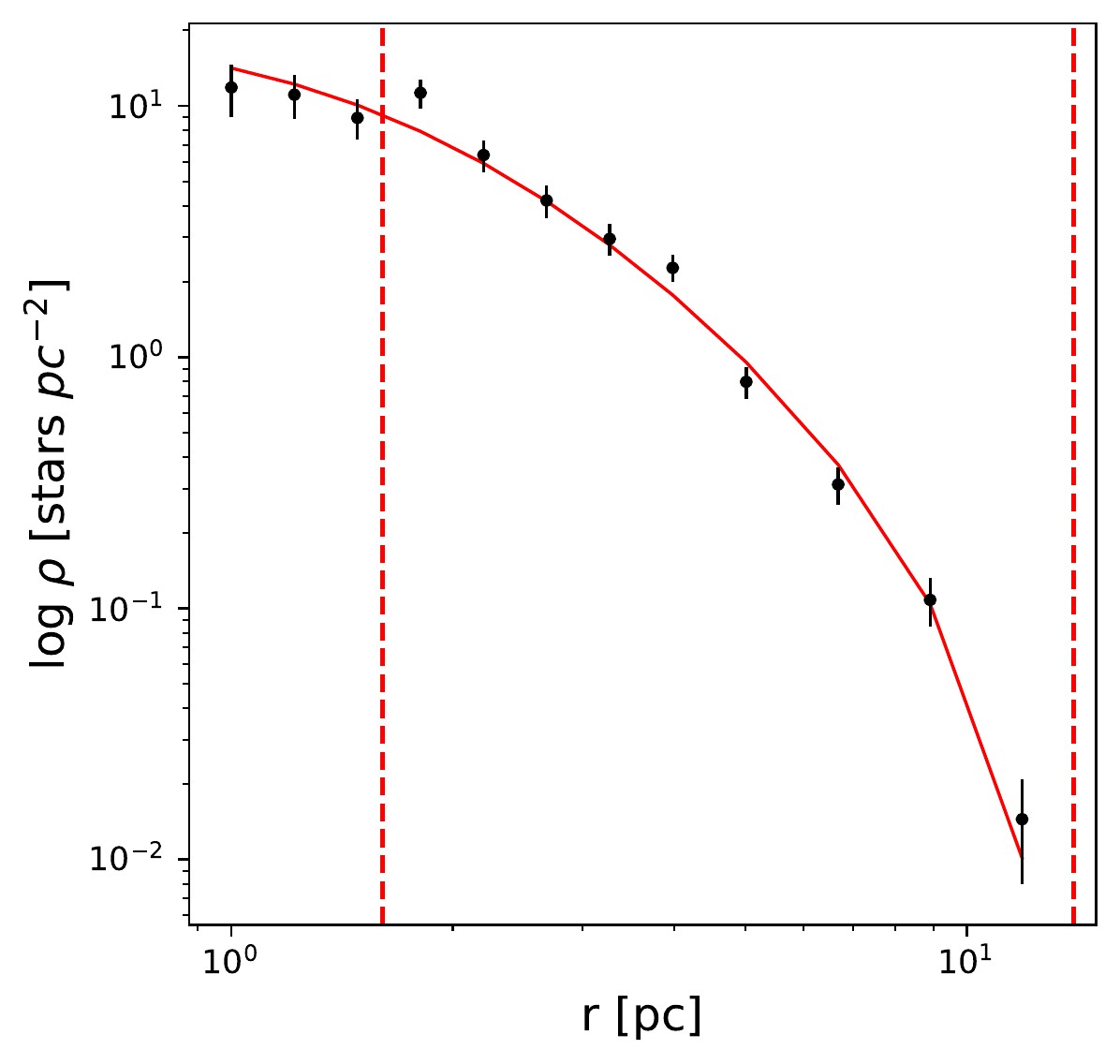}
\caption{RDPs for Pleiades (left panel), Praesepe (middle panel) and Blanco 1 (right panel) OCs. The best-fit King model is shown by the solid red curves. The dashed vertical lines are the $r_c$ and $r_t$ values reported in Table \ref{tab:radius}.}
\label{fig:rdp}
\end{figure*}

The calculation of the RDPs, depicted in Fig. \ref{fig:rdp}, was performed by computing the stellar surface density in concentric rings. The formula utilized was $\rho_i = N_i / \pi (R_{i+1}^2 - R_i^2)$, where $R_i$ and $R_{i+1}$ represent the inner and outer radius, respectively. $N_i$ is the number of stars in the $i_{th}$ ring. The distances of the clusters reported in Table \ref{tab:members} were estimated from a Gaussian fit to the distance histograms, incorporating the corrections outlined by \citet{bailer2015}. To characterize the profiles, the King Model for stellar density defined by Eq. \eqref{equ:king_function} was applied. The MLE method was also implemented to determine the $\rho_0$, $r_c$, and $r_t$ parameters of the model using the log-likelihood function as

\begin{equation}\label{equ:chi_square}
    \ln \mathcal{L}(\rho_0, r_c, r_t) = - \sum_i \left(\frac{\rho_{i} - \rho_{i,King}}{\sigma_{\rho_i}}\right)^2,
\end{equation}

where $\rho_{i}$ is the density computed at the $i_{th}$ ring, $\sigma_{\rho_i}$ is the poissonian uncertainty and $\rho_{i,King}$ is the density predicted by the model. The King model parameters derived from the fit are presented in Table \ref{tab:radius} (see the corner plot of these parameters for the Pleiades in Appendix \ref{appendix:king_stellar}). The density profiles, displayed in Fig. \ref{fig:rdp}, indicate a decrease in density towards low values around 10 pc. This is further supported by the spatial distribution of members, depicted in Fig. \ref{fig:3d}, which show the presence of members at substantial distances from the cluster center.

The Pleiades profile depicted in Fig. \ref{fig:rdp} (left panel), exhibits a gradual decline in the inner region before reaching the core radius. The surface density decreases from $46$ stars pc$^{-2}$ at the center of the cluster to approximately $0.01$ stars pc$^{-2}$ at the tidal radius. The King model provides a good fit to the projected radial distribution of stars because the Pleiades members are mainly within the tidal radius, as demonstrated by the spatial distribution in Fig. \ref{fig:3d} (top panels). The $\rho_0$, $r_c$, and $r_t$ parameters obtained from the fit are presented in Table \ref{tab:radius}. The $\rho_0$ computed in this study is slightly different with the values reported by \citet{pinfield1998} ($32.8$ stars pc$^{-2}$) and close to half of the value reported in \citet{gao2019} ($75.11$ stars pc$^{-2}$). The differences in $\rho_0$ may be due to variations in the number of star members used in these studies. Additionally, both the core and tidal radii computed in this work match with those estimates reported by \citet{lodieu2019} ($2.0$ pc and $11.6$ pc) with a narrow difference of $2.2$ pc in the core radius, and with those in \citet{meingast2021} ($4.4$ pc and $11.8$ pc). They are also approximately $1$ pc different from the values in \citet{gao2019} ($1.27$ pc and $12.3$ pc), which were also obtained through the King model.

\renewcommand{\arraystretch}{1.3}
\begin{table}[h]
\caption{The King model parameters.}
\label{tab:radius}
\centering
\resizebox{\columnwidth}{!}{
\begin{tabular}{@{\extracolsep{4pt}}cccccc}\hline\hline
\multirow{2}{*}{Cluster} & $\rho_0$ & $r_c$ & $r_t$ & \multirow{2}{*}{$\chi^2$} \\  \cline{2-2} \cline{3-4}
 & (stars pc$^{-2}$) & \multicolumn{2}{c}{(pc)} & \\\hline
Pleiades & $46.10^{+3.55}_{-3.27}$ & $2.17^{+0.17}_{-0.16}$ & $11.35^{+0.35}_{-0.21}$ & $0.85$ \\
Praesepe & $29.00^{+2.06}_{-1.85}$ & $2.62^{+0.30}_{-0.26}$ & $12.12^{+0.56}_{-0.60}$ & $0.69$ \\
Blanco 1 & $26.22^{+3.86}_{-3.16}$ & $1.60^{+0.18}_{-0.18}$ & $13.97^{+0.80}_{-0.83}$ & $2.25$ \\\hline
\end{tabular}
}
\tablefoot{The $\rho_0$, $r_c$ and $r_t$ have been found by the best-fit King model with their corresponding $\chi^2$ value.}
\end{table}

The surface density profile for the Praesepe cluster, shown in the middle panels of Fig. \ref{fig:rdp}, displays a similar trend as the Pleiades profile. The King model provides a good match with the observed stellar density up to $10$ pc, however, it fails to accurately reproduce the density beyond this point due to the distribution of members around the tidal radius and beyond. The limitations of the model are further highlighted by the fact that it does not account for tidal debris as noted by \citet{carrera2019}. The King parameters $\rho_0$, $r_c$ and $r_t$ are reported in Table \ref{tab:radius}. The $\rho_0$ obtained in this study is somewhat different by $7.0$ stars pc$^{-2}$ with the value in \citet{holland2000} ($22$ stars pc$^{-2}$). Furthermore, the tidal radii has only $0.62$ pc of variation with that of $11.5$ pc reported by \citet{kraus2007}, and $1.42$ pc with the value of $10.7$ pc in \citet{lodieu2019}.

Regarding of the Blanco 1 cluster, its surface density profile, depicted in the right panel of Fig. \ref{fig:rdp}, exhibits uniformity both before and after the core radius. The King model provides a good representation of the observed density until the tidal radius. Beyond this point, the number of star members decreases to low values, resulting in a reduction of the density in the surroundings and a flattening at around $0.01$ stars pc$^{-2}$. However, the King model cannot reproduce accurately the density beyond the tidal radius by cause of its limitations for elongated clusters like the Blanco 1. The $r_t$ parameter obtained is found to be considerably different from the values reported in \citet{zhang2020} ($10.0$ pc) and \citet{pang2021} ($10.2$ pc) by approximately $3.8$ pc and smaller than the value of $20.0$ pc in \citet{piskunov2007}. 
For the $\rho_0$ we found an increase over the value in \citet{piskunov2007} ($3.2$ stars pc$^{-2}$), but consistent with their core radius $1.5$ pc based on the ASCC-2.5 catalogue. The cause of these discrepancies may be attributed to the methodologies used when determining the parameters and the number of stars included in the {\it Gaia} DR3 data.


\subsection{Spatial distribution}
\label{sec:spatial_distributions}


\begin{figure*}
\centering
\includegraphics[width=\textwidth]{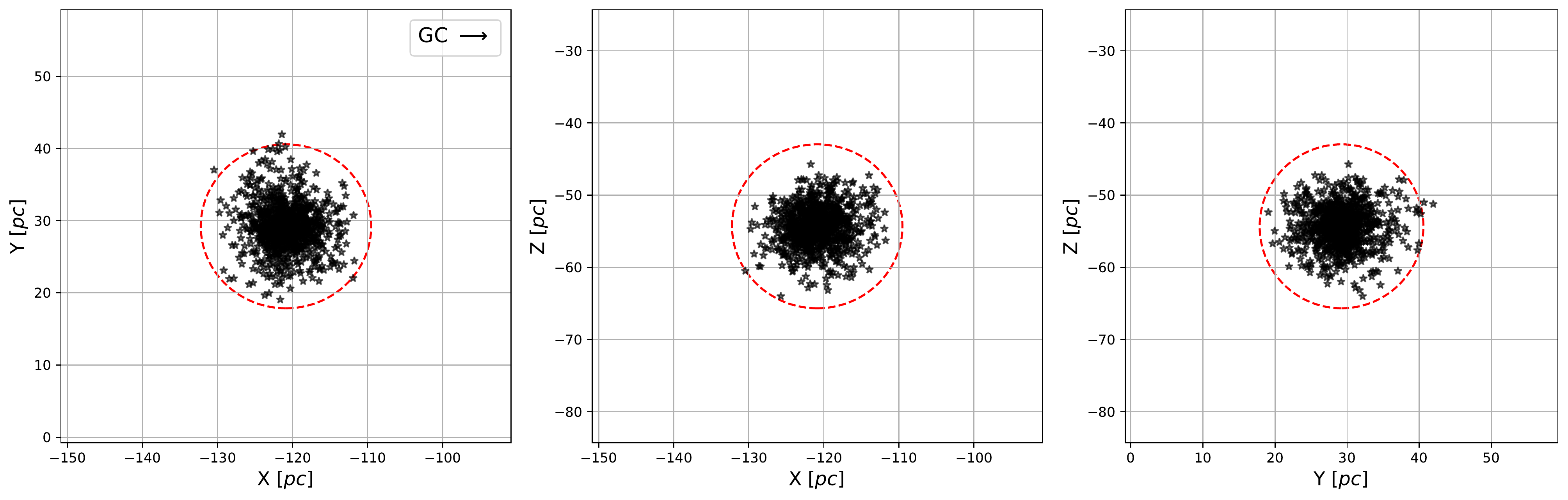}
\includegraphics[width=\textwidth]{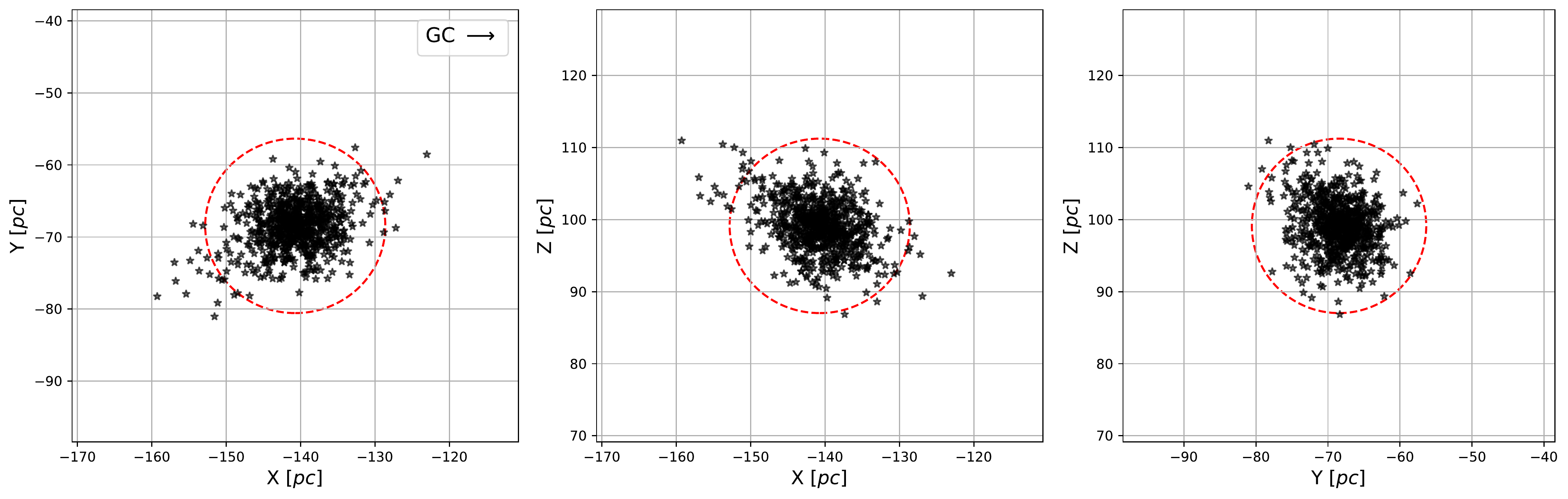}
\includegraphics[width=\textwidth]{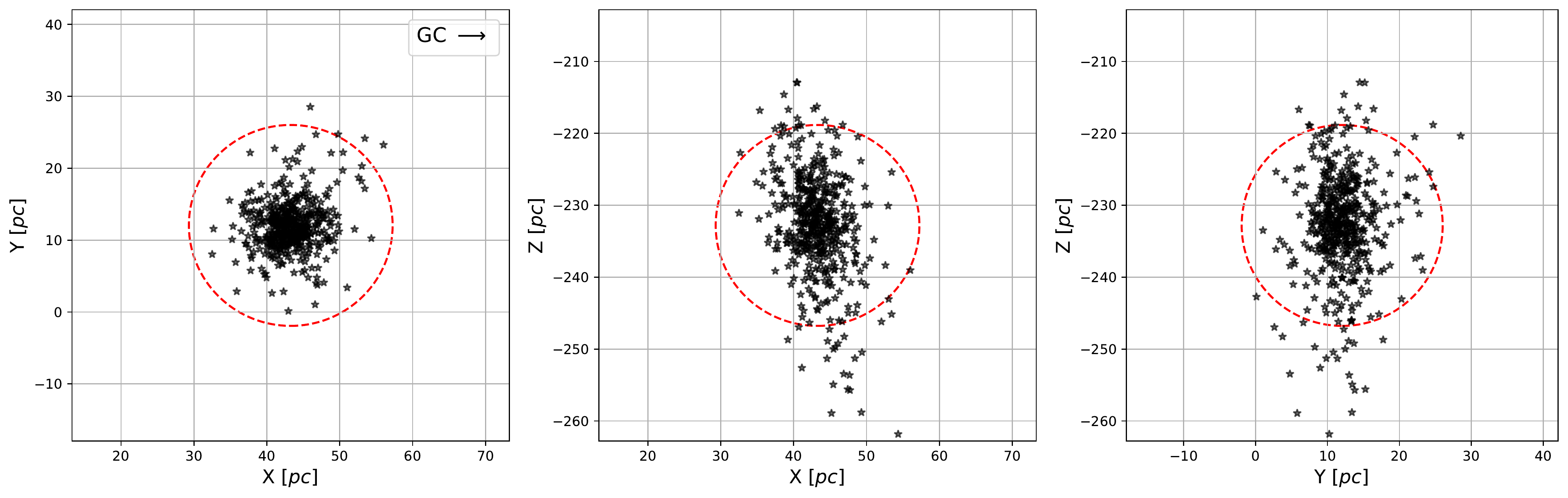}
\caption{Spatial distributions in Galactic cartesian coordinates through the Astropy package \citep{astropy2022} for the Pleiades (top panels), Praesepe (middle panels) and Blanco 1 (bottom panels), after applying the distance corrections. The black arrow points to the Galactic center. The red dashed circles are the tidal radii found in this work reported in Table \ref{tab:radius}.}
\label{fig:3d}
\end{figure*}

Pleiades, Praesepe, and Blanco 1 are widely recognized as some of the most extensively studied star clusters. Due to their proximity to the Sun, these OCs provide a valuable opportunity to evaluate both stellar and dynamic models and to gain insight into the evolution of the Milky Way. From membership results, we have constructed the spatial distributions of these clusters in Galactic cartesian coordinates, which are depicted in Fig. \ref{fig:3d}. The tidal radii reported in Table \ref{tab:radius} for each cluster, are also plotted.

The spatial distribution of the Pleiades cluster is shown in the top panels of Fig. \ref{fig:3d}. Our analysis reveals the presence of $146$ stars within the core radius, $806$ stars between the core and tidal radii, and $6$ stars beyond the tidal radius reaching $13.8$ pc. The latter have high membership probability, indicating that they share similar proper motions with those within the tidal radius. This is in agreement with \citet{gao2019}, who reported members up to $13.8$ pc using probabilities. Our results, however, do not support the detection of a tail-like structure in the (X,Z) plane as reported by \citet{lodieu2019}, but it is also consistent with \citet{meingast2021}, which found more than $80\%$ of its mass within the tidal radius without a prominent tail around it. The non-detection of the tidal tail may be explained by the different methodology used to find the cluster, which restricts the stars in the dimensional space.

Contrary to the Pleiades, the Praesepe members are distributed not only in the core, but also in the surrounding area. Its spatial distribution depicted in Fig. \ref{fig:3d} (middle panels), is characterized by a non-uniform distribution of its members. There are $707$ stars within the tidal radius and $37$ stars beyond it, forming an elongated shape reaching $24$ pc in the (X,Y) and (X,Z) planes, as reported by previous studies, (e.g. \citealt{lodieu2019,roser2019}). This tail-like structure may be a result of the interaction of Praesepe with the Galaxy potential and the stellar evolution because of its old age. Also, the cluster is mass segregated, which allows low-mass stars to be ejected to the outskirts (\citealt{gao2019Praesepe,roser2019}). Subsequently, the bottom panels of Fig. \ref{fig:3d}, shows the spatial distribution of the Blanco 1 cluster, which exhibits a total of $440$ stars within the tidal radius and $48$ stars beyond it. These latter stars are observed to extend up to a distance of $46$ pc and form the most extended tail-like structure among the three clusters analyzed. This elongated shape may be the result of the differential rotation of the Galaxy, as there are no nearby star clusters or molecular clouds that could interact with Blanco 1 \citep{zhang2020}. In addition, \citet{dinnbier2022} has estimated a tilt angle $\beta$ of $-32^{\circ}$ and $22^{\circ}$ for Pleiades and Blanco 1. The latter with a morphological age range from $91$ Myr to $110$ Myr. However, further investigation is necessary to compute this angle in the Praesepe cluster.


\subsection{Age and metallicity}
\label{sec:age_metallicity}

\begin{figure*}[htp]
\centering
\includegraphics[width=0.3\textwidth]{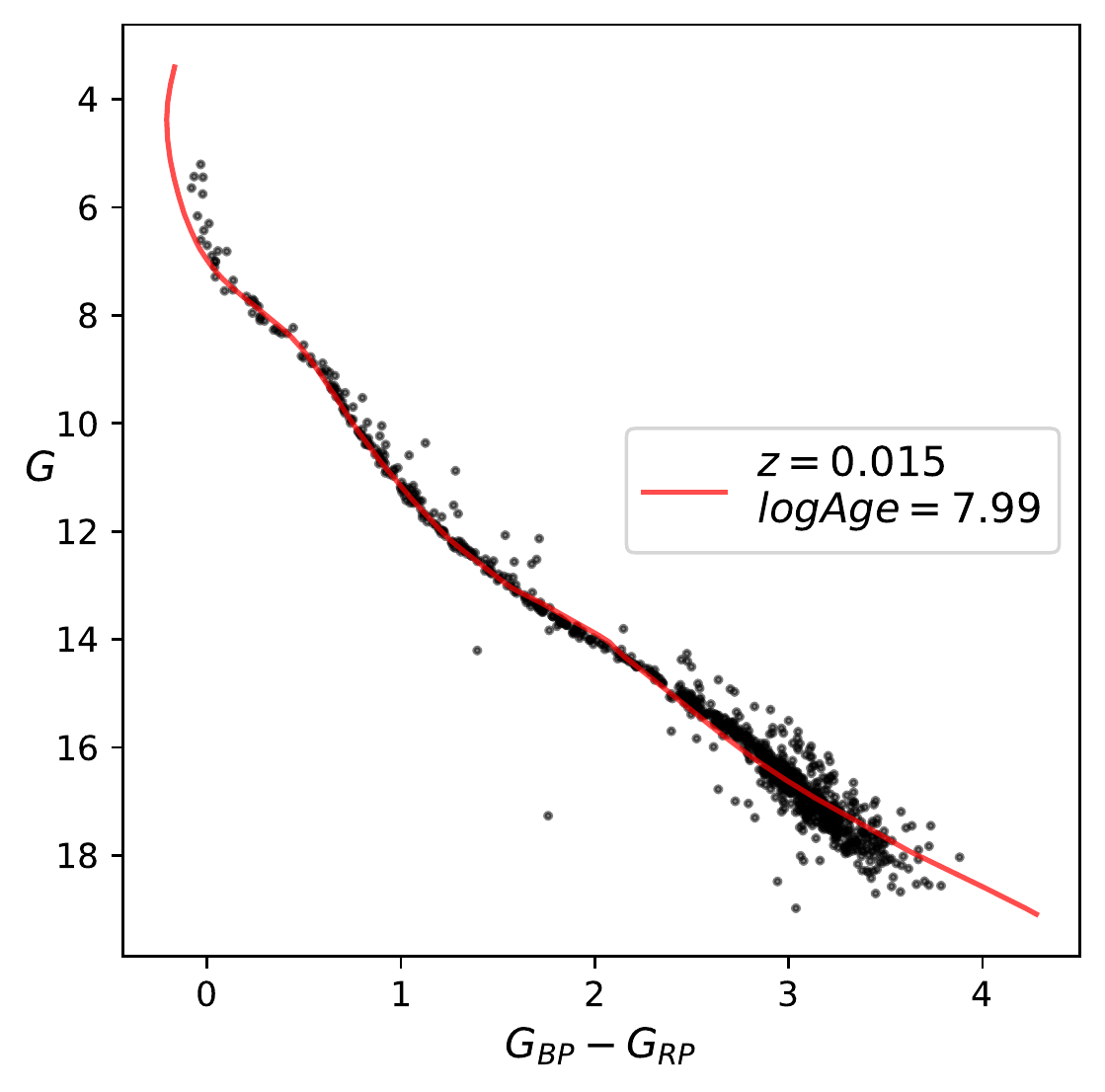}
\includegraphics[width=0.31\textwidth]{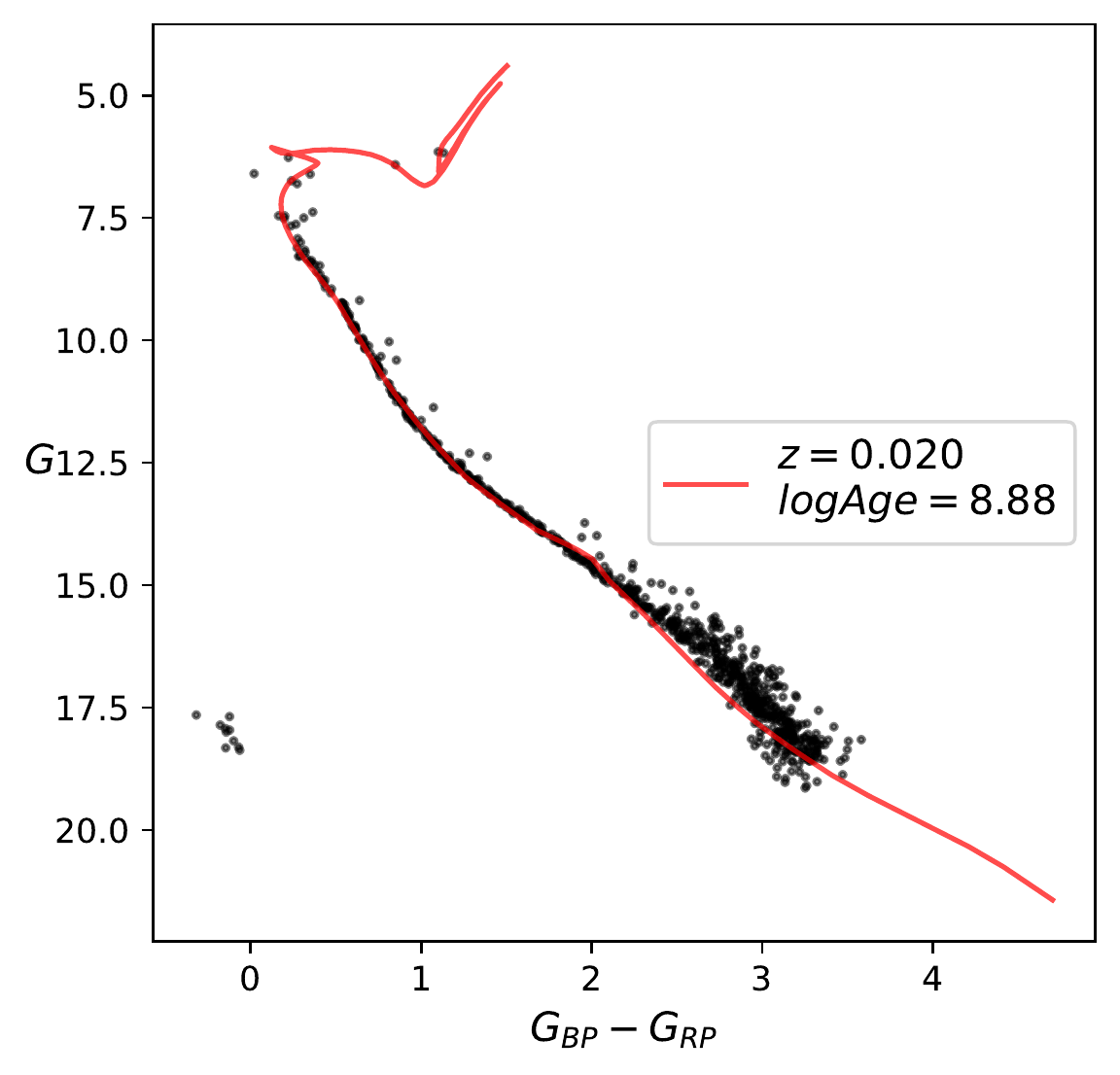}
\includegraphics[width=0.3\textwidth]{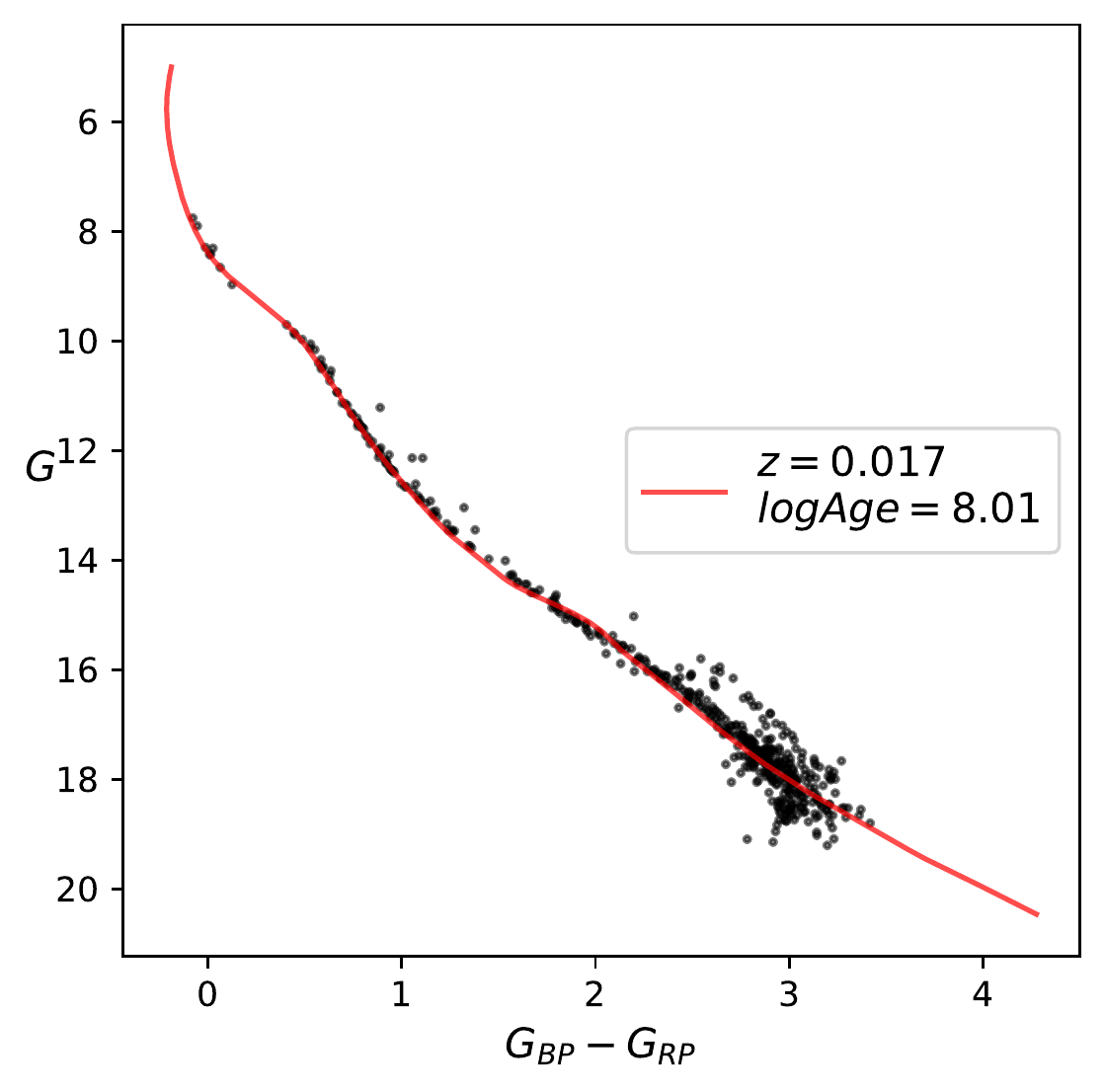}
\caption{CMDs for the Pleiades (left panel), Praesepe (middle panel) and Blanco 1 (right panel) OCs. The best-fit isochrones are indicated with the solid red curves with ages and metallicities estimated using BASE-9 \citep{vonhippel2006,robinson2016}.}\label{fig:isochrones}
\end{figure*}

Two important parameters to understand the dynamic and stellar evolution of star clusters are the age and metallicity. They provide insights on global properties and composition in regions of the Milky Way. Thus, in order to estimate these parameters, we used the PARSEC tracks (\citealt{bressan2012,marigo2013}) and the Bayesian Analysis for Stellar Evolution with nine variables (BASE-9; \citealt{vonhippel2006,robinson2016}) with photometry from the {\it Gaia} DR3 as inputs (see the corner plot of the age and metallicity computed via BASE-9 for the Pleiades in Appendix \ref{appendix:king_stellar}). Because {\it Gaia} does not report magnitude errors, we incorporated uncertainties as $\sigma_{\varpi}$, with a minimum value of $0.02$ mag as done by \citet{kounkel2019}. The best isochrones obtained through BASE-9 and the CMDs are depicted in Fig. \ref{fig:isochrones}.

The estimated age and metallicity values for Pleiades and Blanco 1 star clusters align with previous results reported by \citet{babusiaux2018} and \citet{bossini2019}. In the case of Praesepe, while the computed metallicity concurs with the widely reported value of $0.020$ dex, the estimated age differs from that in \citet{babusiaux2018}. Previous studies have yielded a range of Praesepe age from $662$ Myr to $800$ Myr (e.g. \citealt{mermilliod1981,salaris2019}), however, our computed age is closest to the value of $741.0$ Myr in \citet{gossage2018}.


The star population in the three clusters is mainly governed by main sequence stars, as expected. The brightest reddish stars in the Pleiades CMD depicted in Fig. \ref{fig:isochrones} (left panel), that split of the upper side of the isochrone could be fast rotators \citep{Sun2021}, some of them might be Be stars. Optical spectroscopy is needed to confirm these assumption. Additionally, we did not find the previously confirmed Pleiades white dwarf LB $1497$ \citep{eggen1965}. On the other hand, we found in the Praesepe cluster $10$ out of the $11$ white dwarfs reported by \citet{lodieu2019}. Further studies are needed to establish whether these white dwarfs are part of binary systems. These stars and $3$ in the giant branch with colour around $1.0$ mag were not considered by the isochrone-fit analysis.

In regards to the Blanco 1 cluster, although there are two white dwarf candidates reported by \citet{zhang2020}, we did not detect them in our study. It is possible that the limitations of our method or data may have played a role in the non-detection of these white dwarfs. Further investigation is necessary to confirm their presence or absence.


\section{Summary and conclusions}
\label{sec:conclusions}

We have used two methods to identify the star members in Pleiades, Praesepe, and Blanco 1 OCs using the {\it Gaia} DR3 data. The first method used the MLE and MCMC algorithms to compute the nine parameters of the PM model and their uncertainties. Then, with the estimated parameters, the membership probability could be calculated for each star given its proper motions. A star was selected as a member if its membership probability was greater than or equal to $0.5$. From our analysis we noted that the PM model has difficulty finding the cluster and field centroids when $n_f$ significantly exceeds $n_c$. However, when the cluster kinematics shows mean proper motions greater than those of the background stars, its population can be seen as an overdensity in the VPD. The second method employed the DBSCAN clustering algorithm in a five-dimensional space of positions, proper motions and parallaxes. To find the optimal values for the parameters $\epsilon$ and $minPts$, we followed a similar procedure to the one of \citet{castro-ginard2018}. Our implementation differed by using the median value of $\epsilon$ instead of the minimum value in the $k$-NND. This significantly improved the efficiency of the algorithm in determining the star members in each cluster.

The RDPs were calculated considering Poissonian uncertainties in each ring. To characterize them we fitted the King model implementing the MLE method and the MCMC algorithm to estimate the $\rho_0$, $r_c$, $r_t$ parameters and their uncertainties. The model matched the Pleiades cluster profile because its star members are confined mainly within the tidal radius. For this reason there are no significant differences with previous works in the estimated radii for this cluster. On the contrary, the Praesepe and Blanco 1 clusters have members beyond their tidal radii exhibiting extended structures, and the King model slightly disagrees beyond $10$ pc from the center of the clusters as a result of its limitations. We found substantial increments in the computed tidal radii for these clusters. The most noteworthy is the case of Blanco 1 cluster, whose radius increased compared to those reported by \citet{zhang2020} and \citet{pang2021}, which may be due to the distribution of stars on the outskirts. On the other hand, the computed $\rho_0$ values are larger than those previously estimated with other catalogues because {\it Gaia} has observed these clusters in greater depth. For the Pleaides cluster we found that the stars between the core and tidal radius are a little scattered. This may be product of mass segregation in the cluster \citep{vanleeuwen1980}. Although a tidal tail has been previously reported \citep{lodieu2019}, we could not detect it. For Praesepe and Blanco 1 we have identified elongated shapes and prominent tails containing stars that have high membership probabilities determined from the PM model. In particular, Blanco 1 is the most widespread, reaching a tidal tail almost twice the size of Praesepe's tail. However, we found that only $5\%$ and $10\%$ of the members in  Praesepe and Blanco 1 are on the outskirts of the clusters. In addition, the membership methods used in this work restrict the search region limiting the possibility of observing tails. Further velocity space-based research is needed to recover more stars on the outskirts of these clusters, similar to this performed by \citealt{roser2019,tang2019,jerabkova2021,tarricq2022}.

Although the stellar population in these clusters is mostly dominated by main sequence stars, we found $10$ out of $11$ white dwarfs and $3$ giant stars in Praesepe which had been reported by \citet{lodieu2019}. On the contrary, the Pleiades white dwarf LB $1497$ \citep{eggen1965} was not detected in this work. We also estimated ages and metallicities of the clusters using the BASE-9 method (\citealt{vonhippel2006,robinson2016}) and the PARSEC tracks (\citealt{bressan2012,marigo2013}). The inputs for BASE-9 were the {\it Gaia} photometry values $G$, $G_{BP}$ and $G_{RP}$ of the members classified by the cross-matching. Our age and metallicity results for Pleiades and Blanco 1 are in agreement with those reported by  \citet{babusiaux2018} and \citet{bossini2019}. The result for Praesepe metallicity is in agreement with the obtained by \citet{babusiaux2018}, but its age is close to the value reported by \citet{gossage2018}. 


\begin{acknowledgements}
The authors would like to thank the Vice Presidency of Research \& Creation’s Publication Fund at Universidad de los Andes for its financial support. The authors also acknowledge Prof. Ted von Hippel in helping to set up BASE-9 and to Prof. Beatriz Sabogal for her valuable comments and suggestions to improve this manuscript.  Alejandro Garc\'{i}a-Varela acknowledges financial support given by Fondo de Investigaciones de la Facultad de Ciencias de la Universidad de los Andes, Colombia, through programa de investigación c\'{o}digo INV-2021-128-2295.
This work has made use of data from the European Space Agency (ESA) mission {\it Gaia} (\url{https://www.cosmos.esa.int/gaia}), processed by the {\it Gaia} Data Processing and Analysis Consortium (DPAC, \url{https://www.cosmos.esa.int/web/gaia/dpac/consortium}). Funding for the DPAC has been provided by national institutions, in particular the institutions participating in the {\it Gaia} Multilateral Agreement.
\end{acknowledgements}

%
%

\bibliographystyle{aa}
{\footnotesize\bibliography{bibliography.bib}}


\begin{appendix}

\onecolumn

\section{Gaia query}\label{appendix:gaia_query}

We selected data in a range of $\varpi > 2$ mas. Also, to remove possible artifacts we extracted sources filtering by astrometric errors shown in the query below. 

\begin{small}
\begin{verbatim}
SELECT * FROM gaiadr3.gaia_source WHERE pmra IS NOT NULL AND pmra != 0 AND pmdec IS NOT NULL AND pmdec != 0 
AND ruwe < 1.4 AND phot_g_mean_flux_over_error > 10 AND phot_rp_mean_flux_over_error > 10 
AND phot_bp_mean_flux_over_error > 10 AND visibility_periods_used > 8 AND astrometric_excess_noise < 1 AND 
parallax_over_error > 10 AND parallax IS NOT NULL AND parallax > 2
\end{verbatim}
\end{small}


\section{Uncertainty estimation of the nine parameters in the PM model}
\label{appendix:mcmc}

Corner plot of the sampling of the posterior probability function using the MCMC algorithm for the PM model in the Pleiades region. The implementation uses $100$ walkers and $5000$ iterations, and converges to a good agreement with the MLE method result. The upper and lower uncertainty in each parameter corresponds to the $16_{th}$ and $84_{th}$ percentiles of the samples, respectively. 

\begin{figure*}[h]
    \centering
    \includegraphics[scale=0.32]{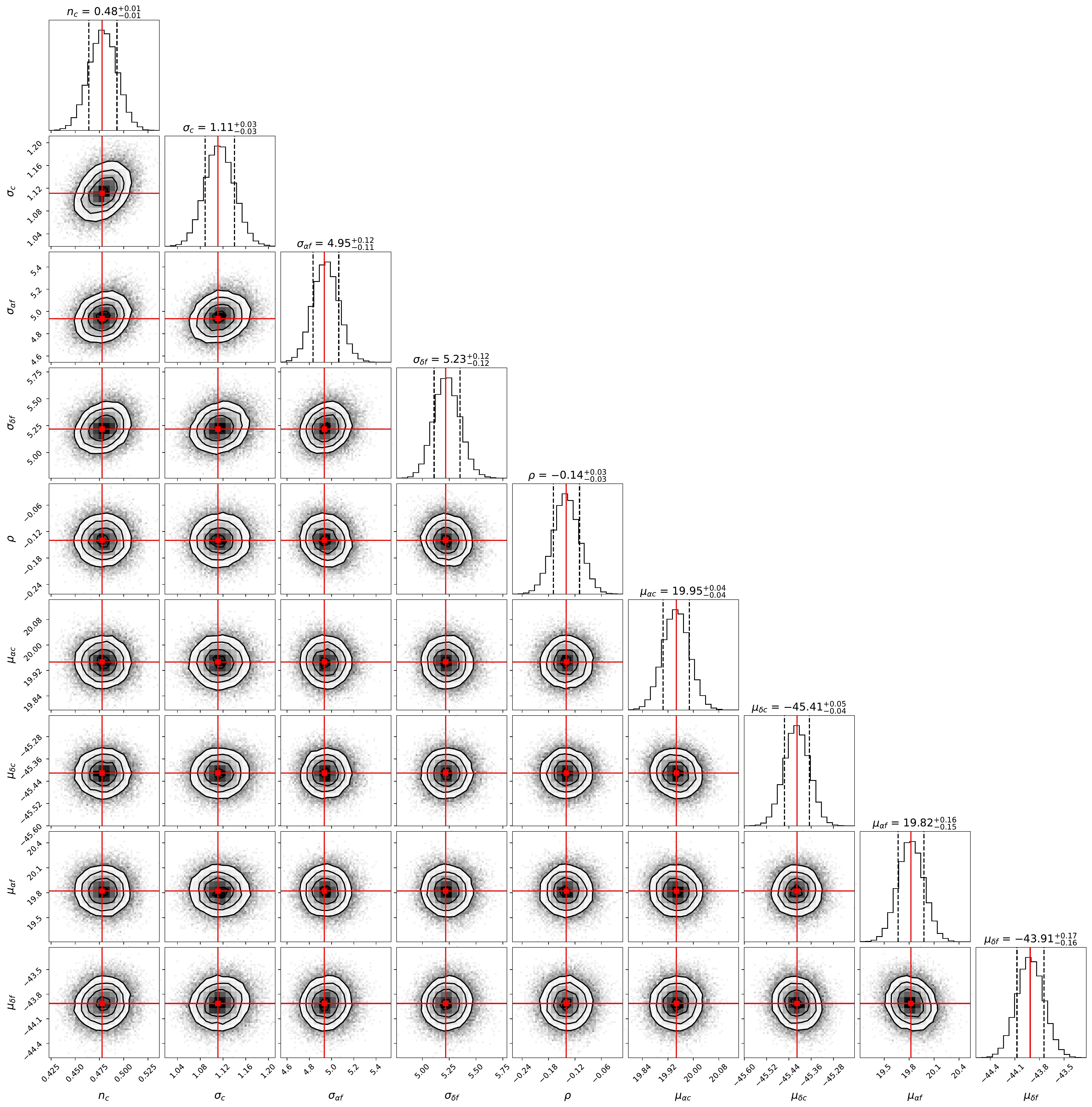}
    \caption{Corner plot of the posterior probability density function of the PM model for the Pleiades region.}
    \label{fig:corner}
\end{figure*}


\section{Estimation of the King and stellar parameters}
\label{appendix:king_stellar}

Corner plots of the King and stellar parameters sampling of the posterior probability functions using the MCMC algorithm for the Pleiades cluster. The upper and lower uncertainty in each parameter corresponds to the $16_{th}$ and $84_{th}$ percentiles of the samples, respectively.

\begin{figure}[H]
    \centering
    \includegraphics[scale=0.5]{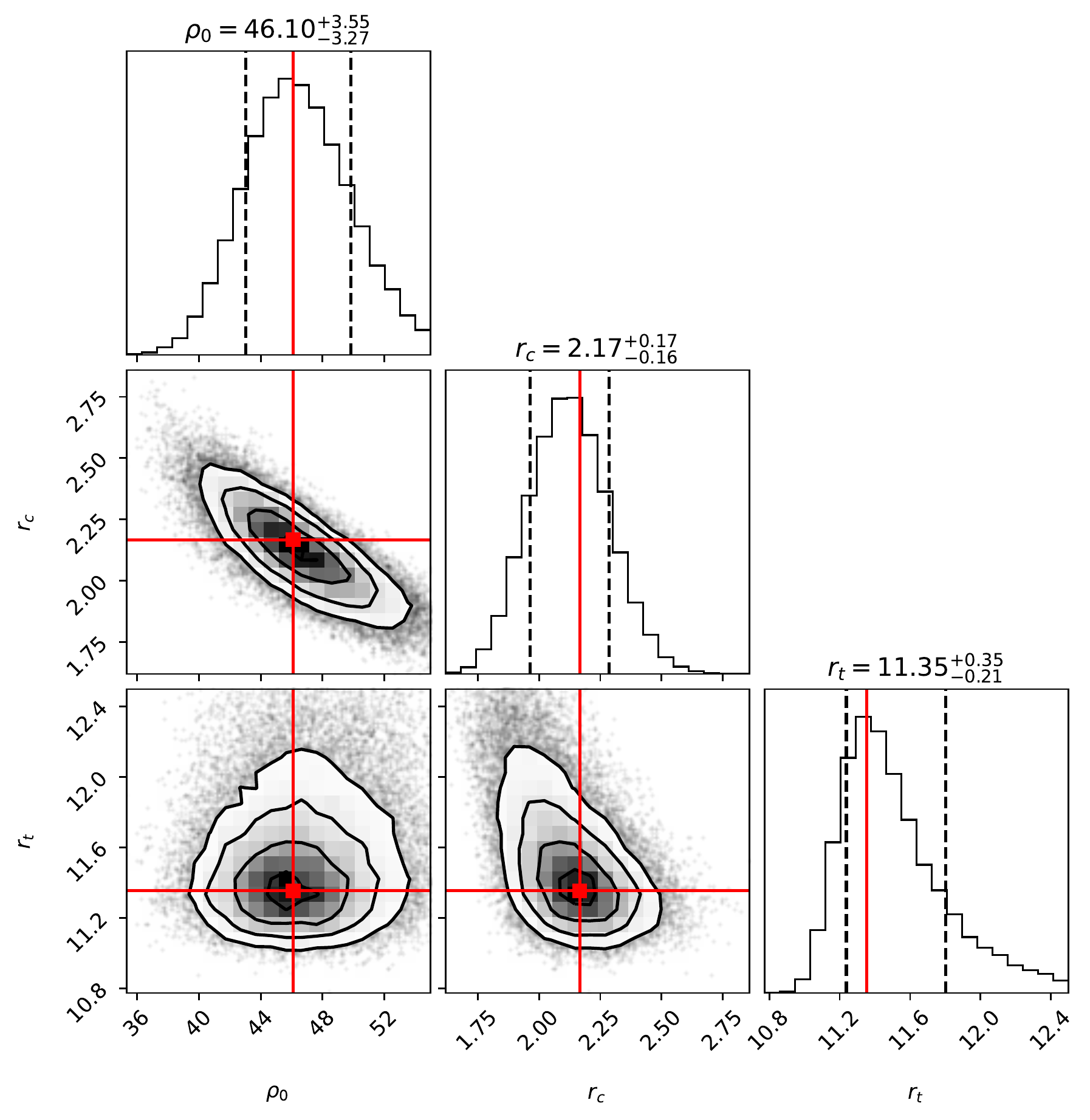}
    \includegraphics[scale=0.7]{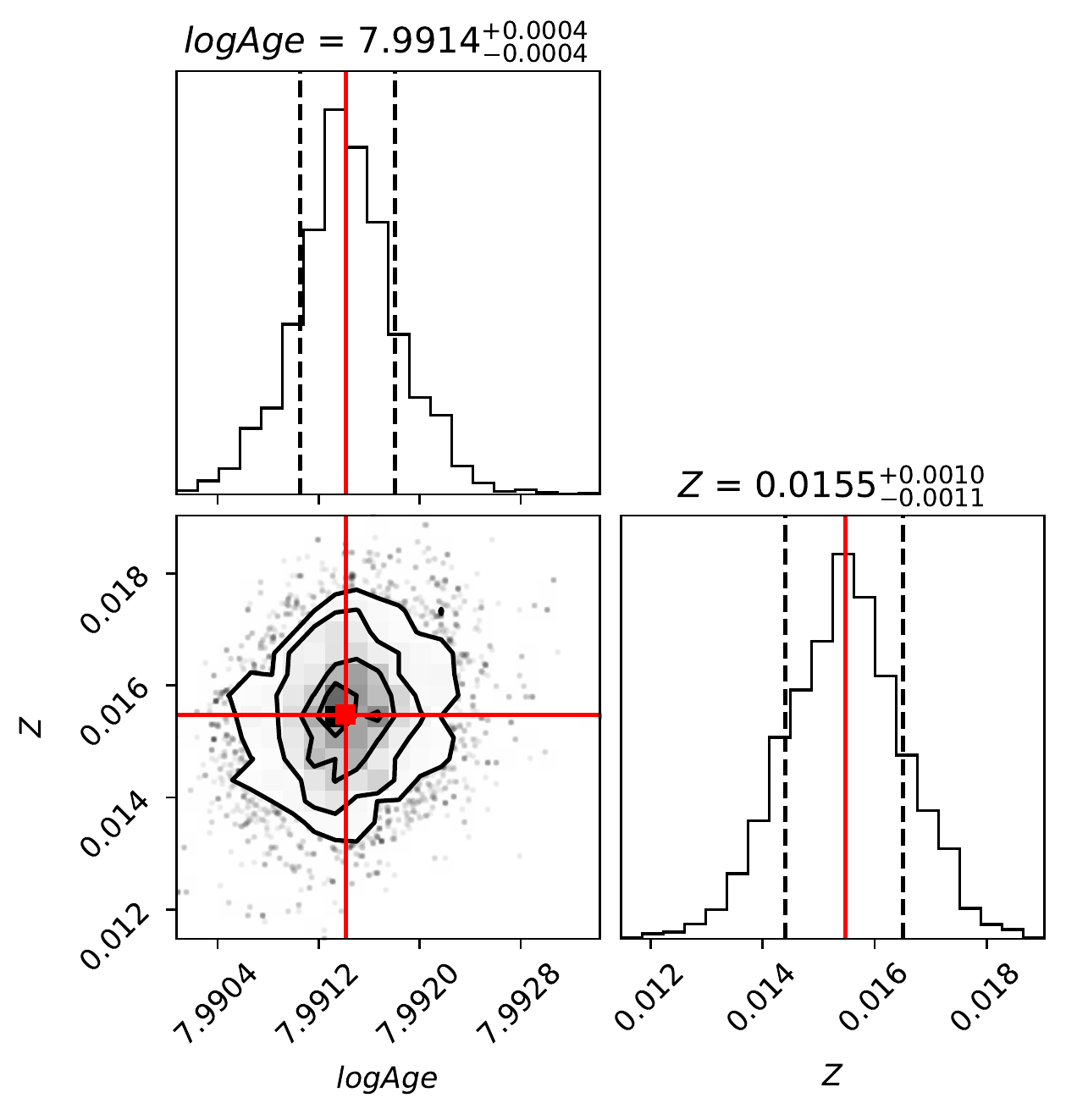}
    \caption{Corner plots of the estimations of the King parameters (left panel) and stellar parameters using BASE-9 (right panel) for the Pleiades cluster.}
    \label{fig:corner_king_stellar}
\end{figure}

\end{appendix}


\end{document}